\documentclass[lettersize,journal]{IEEEtran}
\usepackage{amsmath,amsfonts}
\usepackage{algorithmic}
\usepackage{array}
\usepackage[caption=false,font=normalsize,labelfont=sf,textfont=sf]{subfig}
\usepackage{textcomp}
\usepackage{stfloats}
\usepackage{url}
\usepackage{verbatim}
\usepackage{graphicx}
\usepackage{booktabs}
\usepackage{marvosym}
\def\BibTeX{{\rm B\kern-.05em{\sc i\kern-.025em b}\kern-.08em
    T\kern-.1667em\lower.7ex\hbox{E}\kern-.125emX}}
\usepackage{balance}
\begin{document}
\title{Tensorformer: Normalized Matrix Attention Transformer for High-quality Point Cloud Reconstruction}
\author{Hui Tian\textsuperscript{1}, Zheng Qin\textsuperscript{1}, Renjiao Yi\textsuperscript{1}, Chenyang Zhu\textsuperscript{1}, Kai Xu\textsuperscript{1 \Letter}
\thanks{1, National University of Defense Technology}}

\markboth{Journal of \LaTeX\ Class Files,~Vol.~18, No.~9, September~2020}%
{}

\maketitle

\begin{figure*} 
\centering
  \includegraphics[width=2\columnwidth]{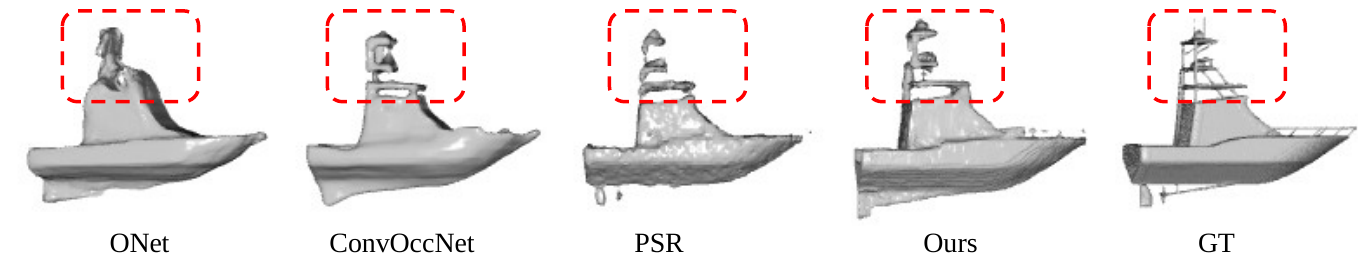}
  \caption{Comparison with baseline methods. Due to the powerful local feature representation, Tensorformer can better recover thin structures than alternatives, e.g., PSR~\cite{2013Poisson}. Our global shape is more faithful to the ground-truth mesh than ConvOccNet\cite{2020Convolutional} and ONet\cite{2019Occupancy}}
  \label{fig:teaser}
\end{figure*}

\begin{abstract}
 Surface reconstruction from raw point clouds has been studied for decades in the computer graphics community, which is highly demanded by modeling and rendering applications nowadays. Classic solutions, such as Poisson surface reconstruction, require point normals as extra input to perform reasonable results. 
Modern transformer-based methods can work without normals, while the results are less fine-grained due to limited encoding performance in local fusion from discrete points. 
We introduce a novel normalized matrix attention transformer (Tensorformer) to perform high-quality reconstruction. The proposed \emph{matrix attention} allows for simultaneous point-wise and channel-wise message passing, while the previous vector attention loses neighbor point information across different channels. It brings more degree of freedom in feature learning and thus facilitates better modeling of local geometries. Our method achieves state-of-the-art on two commonly used datasets, ShapeNetCore and ABC, and attains 4\% improvements on IOU on ShapeNet. Our implementation will be released upon acceptance.
\end{abstract}
 

\section{Introduction}
\label{intro}

Surface reconstruction from point clouds is important yet challenging in many computer vision and graphics applications, such as geometry modeling and AR rendering. Before deep learning methods arose, traditional methods such as Poisson Surface Reconstruction (PSR)\cite{2013Poisson} have been proposed to address this challenge. The PSR can generate a smooth surface with per-point normal by solving a Poisson equation of indicator function. However, per-point normal is always difficult to acquire in real applications like autonomous driving. Meanwhile, the absence of machine learning modules makes these traditional methods lack the ability to recover local details while the input points are sparse or noisy.

With rapid development of deep learning, some works try to utilize the strong prediction ability of the neural network to improve the reconstruction quality. Most CNN-based works\cite{2019DeepSDF, 2019Occupancy, 2020Convolutional, deepmls} define the surface reconstruction as a classification or regression problem. Specifically, the neural network should predict the SDF or occupancy value of all points in the whole space or a bounding box according to the point cloud. DeepSDF\cite{2019DeepSDF} extracts the global feature via PointNet\cite{2017PointNet} of the whole point cloud and combines feature and query point to predict the occupancy. Convolutional Occupancy Networks\cite{2020Convolutional} casts the per-point feature extracted via PointNet++\cite{2017PointNet2} to fixed grid point and uses 3D convolution to enhance the feature. All the above methods employ convolution and its variation to enhance the extracted features. The good per-point feature is critical for reconstructing high-quality surfaces.

Inspired by the recent success of transformers on natural language processing~\cite{2017Attention} and image understanding~\cite{vit}, a lot of works have attempted to extend transformer to point cloud feature encoding~\cite{engel2020pointtrans,zhao2021pointtrans} which has achieved good performance for segmentation and classification. The transformer is particularly suited for point cloud processing since self-attention, which is the center of a transformer, is essentially a set operator~\cite{zhao2021pointtrans}. However, current Scaled Dot-product Attention\cite{2017Attention} and Vector Attention\cite{2020Vector} only aggregate the feature of the same channel, neglecting the importance of feature aggregation between different channels, which is not enough for local detailed geometry recovery from a sparse point cloud.

\begin{figure}
\centering
  \includegraphics[width=\columnwidth]{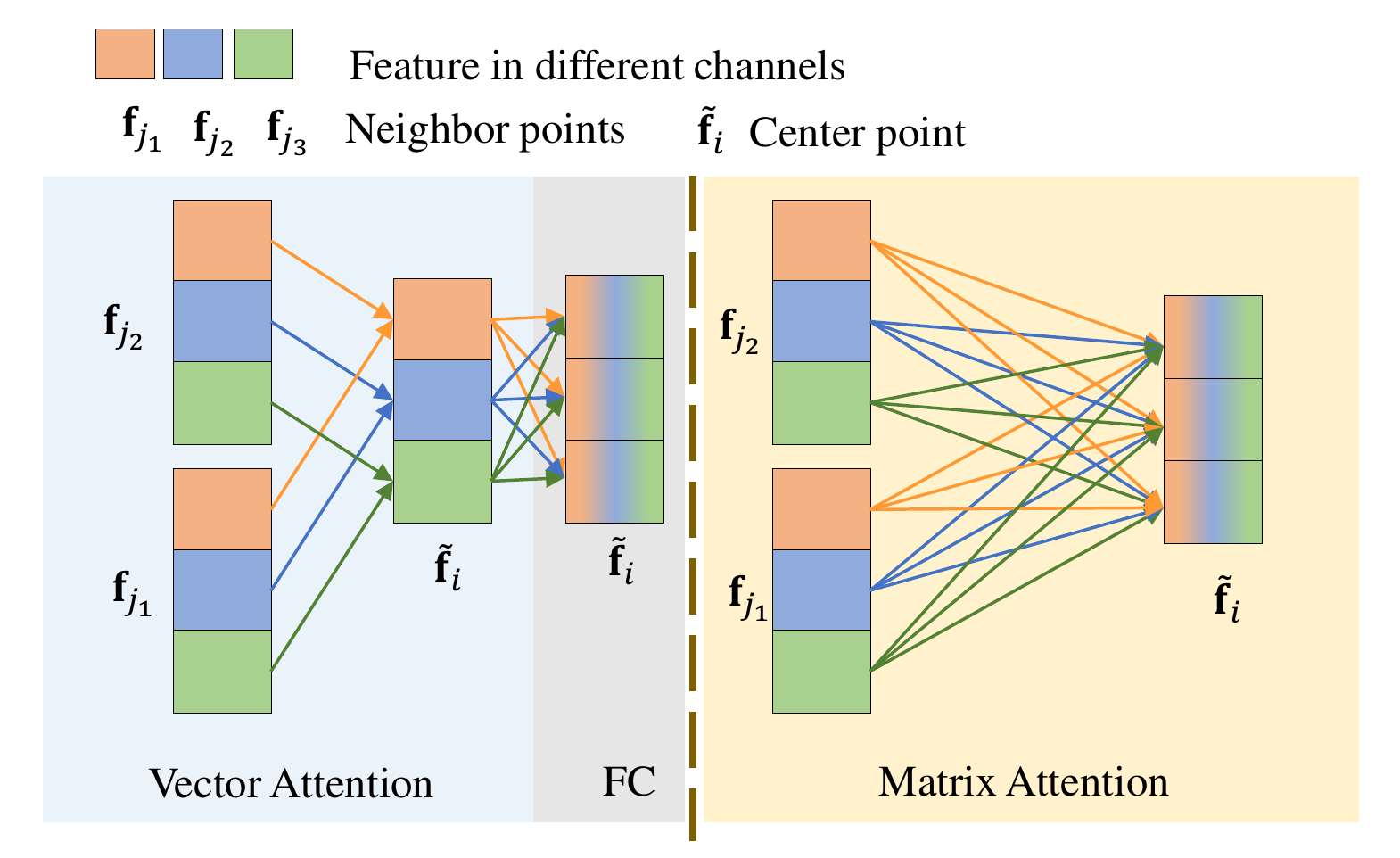}
  \caption{The difference between Vector Attention + FC and Matrix Attention. We ignore the attention weights and focus on the information flow here. Vector Attention + FC first performs a channel-wise feature fusion. The features are then weighted and fused across channels via FC, illustrated with $K\times C + C^2$ arrows. Matrix Attention can make feature fusion across channels and neighbors simultaneously, illustrated with $K\times C^2$ arrows. A arrow represents one scalar value flow in feature fusion.
}
  \label{fig:diff}
\end{figure}

We propose \emph{Matrix Attention}, a novel self-attention module for encoding the geometry of 3D point clouds. In matrix attention, features of different neighboring points and different channels of each point are simultaneously weighted and fused, leading to more degrees of freedom. Such joint point-wise and channel-wise attention help better capture the local geometry, in analogy to the joint set and coordinate embedding of discrete differential operators on 2-manifold in 3D space~\cite{meyer2003discrete}. Fig. \ref{fig:diff} depicts the difference between Vector Attention + FC and Matrix Attention. Our experiment in table~\ref{tab:ab} demonstrates the superiority of Matrix Attention among other attention designs in the reconstruction task.

Existing self-attention modules usually employ a softmax normalization for outputting the final attention weights. Softmax yields a high variance of attention weights, which adversely affects local geometry encoding. We will provide a theoretic analysis of the method in Sec.~\ref{sec:method}. To further release the power of our attention design, we found that a simplistic weight normalization results in a powerful transformer for geometry learning. With the designs above, we name our new point transformer, which is short for \emph{Tensorformer}. Fig. \ref{fig:teaser} demonstrates that the proposed method performs better for local geometry details preserving and generating more clear surfaces than other state-of-the-art alternatives. In summary, the contributions of this paper include the following: 
\begin{itemize}
\item We propose matrix attention which can simultaneously integrate features from different channels and neighbors, significantly improving the point feature expression ability for reconstruction.
\item We propose linear normalization, which can prevent the gradient vanish and back-propagate the gradient to almost all the input elements rather than the most significant ones.
\item We establish a network with Tensorformer and achieve SOTA on ShapeNetCore and ABC(001) on point cloud reconstruction. Our method performs an average improvement of 5\% on IoU from the $13$ categories of the ShapeNetCore dataset.
\end{itemize}

\section{Related Work}

\paragraph{3D point representation learning} Among 3D shape representation methods, point cloud might be the easiest one to make use of deep neural network. Naturally, many methods have emerged to handle raw point cloud directly. PointNet\cite{2017PointNet,zhang2017airport, li2017fast} are the pioneers and a milestone in point cloud processing. it use point-wise MLP to extract feature. PointNet++\cite{2017PointNet2} follows it and add local feature aggregation module. Then, lots of point convolution networks appear. Pointwise CNN\cite{song2021pointwise} locates the kernel weights with voxel bins. SpiderCNN\cite{2018SpiderCNN} tries to define the kernel weight as a family of polynomial functions. Flex-convolution\cite{groh2018flex} uses a linear function to model the kernel. PointConv\cite{2019PointConv} adopts similar way with flex-convolution and adds adaptive density module. In KPConv\cite{thomas2019kpconv}, The convolution weights are located in Euclidean space by kernel points, and applied to the input points close to them. Similarly, our method tries to extract more expressive point features, but with a new self-attention module.

\begin{figure*} 
\centering
  \includegraphics[width=2.1\columnwidth]{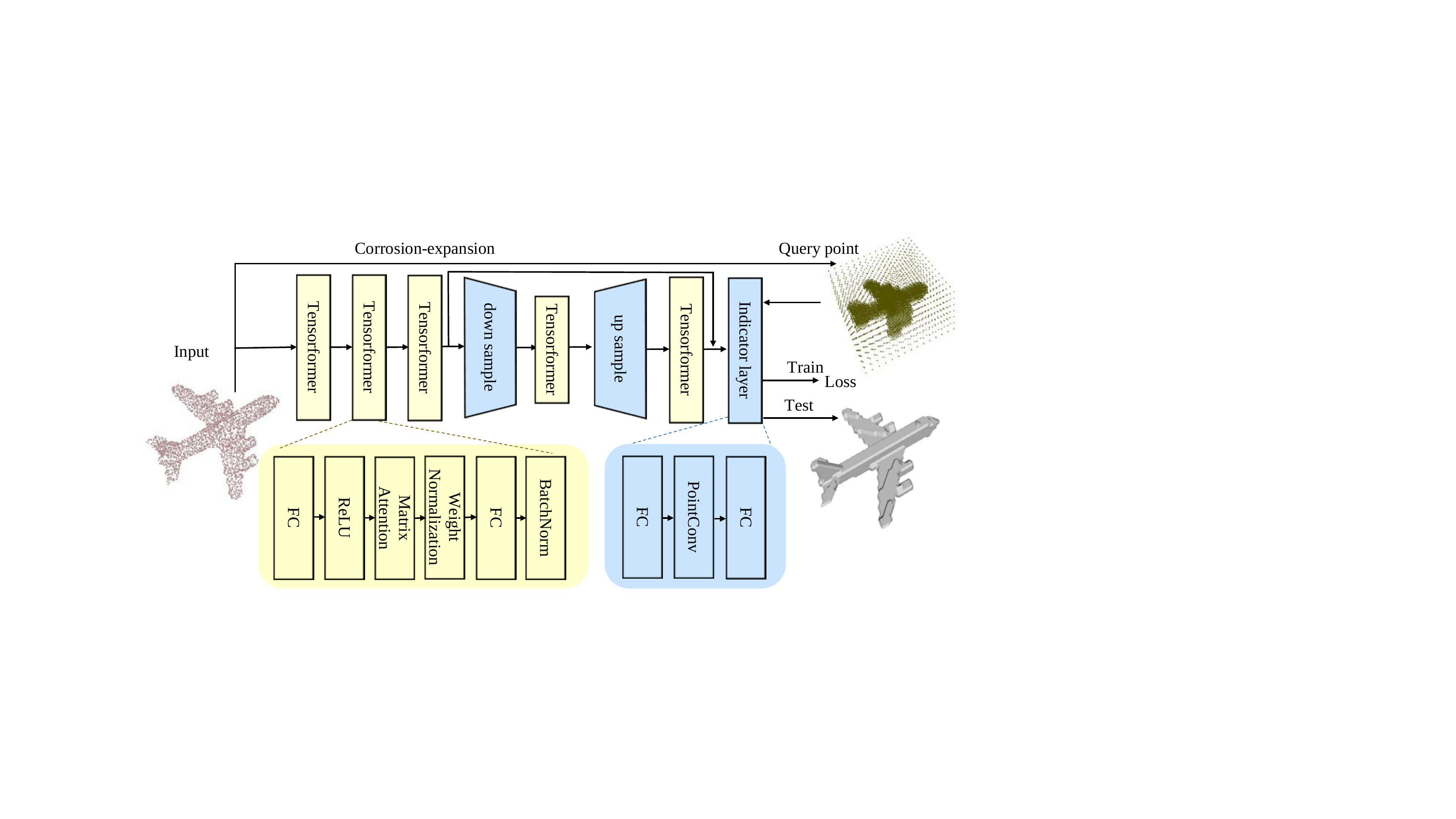}
  \caption{The network architecture for point cloud reconstruction.}
  \label{fig:network}
\end{figure*}

\paragraph{Transformer} Transformer is a convolution-free method, using attention weight to substitute convolution weight in feature aggregation. At the beginning, it\cite{2017Attention} is applied in NLP to capture long-term word tokens relation and achieves excellent performance. Later, the computer vision community notices the transformer's potential for vision task. Large number of works has emerged in last 2 years related to computer vision. Vision Transformer(ViT)\cite{vit} applies transformer in image classification by partitioning the image into fixed patches and using transformer to pass information between different patch. DETR\cite{detr} uses a transformer encoder-decoder architecture to predict bounding box. Pyramid Vision Transformer\cite{wang2021pyramid} extracts different resolution feature by transformer and establishes an unified backbone for varied vision task. Swin Transformer\cite{swin} uses shifted window to make information flow in different patch and also establishes a backbone for varied vision task. 

In 3D vision, Point Transformer\cite{zhao2021pointtrans, engel2020pointtrans} and Point Cloud Transformer\cite{pct} almost appear at the same time. They apply transformer to point cloud processing, and achieve state of the art on classification and segmentation task. Due to the calculation complexity, some other works try to reduce the computation cost by modifying the attention mechanism.  \cite{beltagy2020longformer}, \cite{child2019generating} explore selective
or sparse attention to previous layer tokens which updating
each next layer token.  All the methods mentioned above use dot-product scalar attention or vector attention.

\paragraph{3D point cloud reconstruction} Point cloud reconstruction has been a long-standing task in graphics and vision. The most important traditional method is Poisson surface Reconstruction~\cite{2013Poisson} and ball-pivoting reconstruction~\cite{ballpivoting}. The former method classifies the query point according to the Poisson indicator function. The latter construct the continuous surface by making up a ball rolling on the points. Those two methods can reconstruct pretty good surface, however the methods need per-point normal as input.

Deep methods can be divided into point-based and voxel-based. Voxel-based methods first convert the point cloud into voxels and then apply volume convolution to get the feature of every voxel, just like o-net\cite{2019Occupancy} and O-CNN\cite{o-cnn}. Point-based methods can be divided into local methods and global methods. As the name implies, when giving a query point, global methods classify it according to the whole shape information, local methods classify the query point according to the neighbor points of it. Representative global methods are DeepSDF\cite{2019DeepSDF}, BSP-Net\cite{chen2020bsp} and Predictive Context Priors\cite{2022contextprior}. The routine of those methods are extracting the feature code of the whole shape and then recovering the surface from the code. Representative local methods are SSR-Net\cite{2020SSRNet}, DeepMLS\cite{deepmls}, POCO\cite{2022POCO}, Dual Octree\cite{dualoctree} and Dynamic Code\cite{2022dynamicCode}. Those local methods typically use the help of octree. SSR-Net firstly extracts point feature, then maps the neighbor-hood points feature to octants, finally classifies the octants. DeepMLS tries to predict the normal and radius of every point, then classify the query points according to moving least-squares equation. Further more, except for local methods and global methods, there are some methods try to combine global and local information. The most representative method is Points2Surf\cite{2020Points2Surf}. It tries to regress the absolute SDF value according to local information and classify the sign according to global information.

Another type of deep point cloud reconstruction method tries to encode the 3D shapes with a neural network directly, the input of the network is coordinate position, the output is SDF(signed distance function) or UDF(unsigned distance function). In test period, these methods need training as well. Representative methods are Neural Pull\cite{2020npull} and On-surface Prior\cite{onsurfaceprior}.

\section{Method}
\label{sec:method}

In this section, we propose a novel attention-based network for point cloud reconstruction. An overview of our architecture is illustrated in Fig.~\ref{fig:network}. For a given point cloud, we first design a Normalized Matrix Attention Transformer (Tensorformer) to learn point-wise features (Sec.~\ref{sec:Tensorformer}). We then sample a set of query points and predict their occupancy value with a Indicator Module (Sec.~\ref{sec:indicator}). At last, the final mesh is constructed by applying Marching Cube \cite{lorensen1987marching} on the query points. 

\begin{figure*}
\centering
  \includegraphics[width=2\columnwidth]{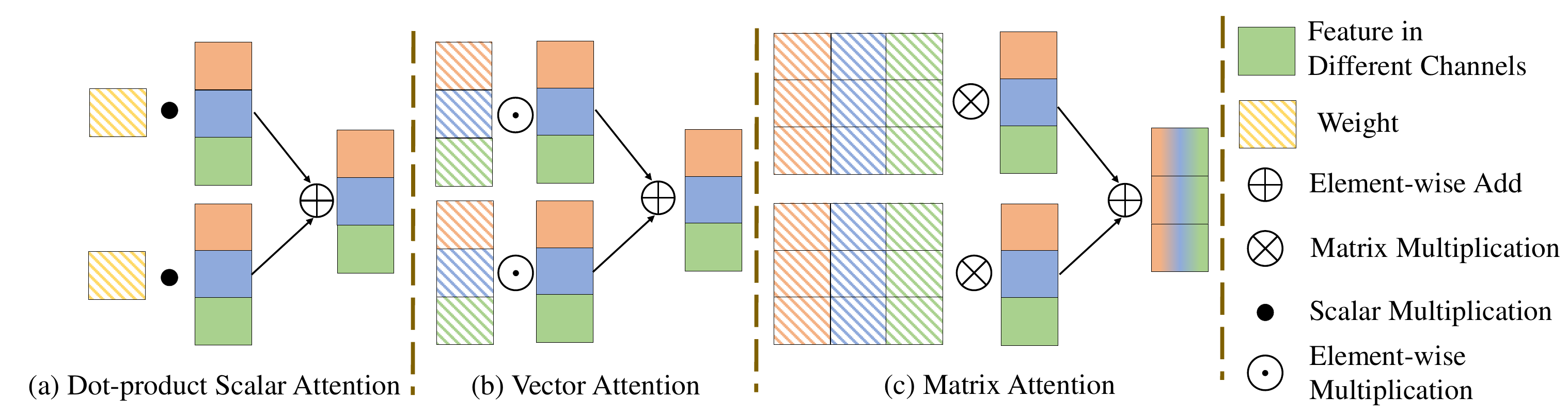}
  \caption{The difference between scaled dot-product attention, vector attention and matrix attention. For simplicity, let us assume the point being considered has only two neighboring points. (a) is scaled dot-product attention. (b) is vector attention, which gives each channel a different attention to improve the expression ability. (c) is matrix attention, where the features can be fused across channels and neighbors simultaneously. 
}
  \label{fig:nv}
\end{figure*}

\subsection{Normalized Matrix Attention Transformer}
\label{sec:Tensorformer}



Inspired by the recent success of point transformers \cite{engel2020pointtrans, zhao2021pointtrans}, we utilize an attention-based structure to learn point-wise features for the point cloud by aggregating features from a local patch. Formally, given an anchor point $\mathbf{p}_i$ with its features $\mathbf{f}_i$ and its $k$ nearest neighbors $\{ \mathbf{p}_{n_1}, ..., \mathbf{p}_{n_k} \}$ with their features $\{ \mathbf{f}_{n_1}, ..., \mathbf{f}_{n_k} \}$, our goal is to compute the output features $\mathbf{z}_i$ of $\mathbf{p}_i$. And the computation of a local patch attention can be formulated as:
\begin{equation}
\label{eq:attention}
\mathbf{z}_i = \sum_{j=1}^{k}\mathcal{A}(\mathtt{Norm}_j(\mathcal{W}(\mathbf{f}_i, \mathbf{f}_{n_1}),..., \mathcal{W}(\mathbf{f}_i, \mathbf{f}_{n_k})), \mathbf{f}_{n_j}),
\end{equation}
where $\mathcal{W}$ is a attention weight function, $\mathtt{Norm}$ is a normalization function, and $\mathcal{A}$ is a feature aggregation function.
We assume the feature dimension is $d$ and omit the query, key, value projections here for simplicity.
Feature aggregation could be conducted from two dimensions, i.e., the feature dimension (across feature channels) and the spatial dimension (across neighbor points). Previous attention mechanisms can be classified into two categories: scaled dot-product attention \cite{2017Attention,zhao2021pointtrans} and vector attention \cite{2020Vector}. However, these two attentions can only aggregate the features in the spatial dimension but ignore the feature dimension, which harms the expressive ability. Next, we will first briefly review the two attentions and then present a new attention named matrix attention which realizes feature aggregation simultaneously in both dimensions. 


\paragraph{Scaled Dot-product Attention} In scaled dot-product attention, an scalar attention weight for each neighbor point $\mathbf{p}_{n_j}$ is computed as the inner product between the features of the anchor point $\mathbf{f}_i$ and the neighbor point $\mathbf{f}_{n_j}$:
\begin{equation}
\label{eqn:dot-product-att}
  \mathbf{z}_i = \sum_{j=1}^{k}\mathtt{Norm}_j(\mathbf{f}_i^{T}\mathbf{f}_{n_1},...,\mathbf{f}_i^{T}\mathbf{f}_{n_k})\mathbf{f}_{n_j},
\end{equation}
where $\mathtt{Norm}_j$ is usually the j-th element of softmax operation along the spatial dimension to weigh the importance of different neighbor points.
As shown in Eq.~\ref{eqn:dot-product-att}, the aggregation function $\mathcal{A}$ here is simply a channel-wise weighted sum of the neighbor point features and thus no inter-channel feature exchange is conducted.

\paragraph{Vector Attention} In vector attention, an attention weight vector is computed for each neighbor point $\mathbf{p}_{n_j}$ from its feature differences $\mathbf{f}_i - \mathbf{f}_{n_j}$ between the anchor point:
\begin{equation}
\label{eqn:vector-att}
  \mathbf{z}_i = \sum_{j=1}^{k}\mathtt{Norm}_j(\boldsymbol{\Phi}(\mathbf{f}_i-\mathbf{f}_{n_0}),...,\boldsymbol{\Phi}(\mathbf{f}_i-\mathbf{f}_{n_k})) \odot \mathbf{f}_{n_j},
\end{equation}
where $\boldsymbol{\Phi}: \mathbb{R}^{d} \to \mathbb{R}^{d}$ is a learnable function, $\odot$ represents Hadamard product, and $\mathtt{Norm}$ is usually a softmax operation along the spatial dimension. Compared to scaled dot-product attention, vector attention gives each channel an independent attention weight, thus increasing the expression ability. However, as in scaled dot-product attention, there is still no information communication across the feature channels in the feature aggregation process.

\paragraph{Matrix Attention}
To address this issue, we propose a novel attention mechanism named \emph{Matrix Attention} which enables feature aggregation in both spatial and feature dimensions simultaneously.
Our core idea is to extend the attention weights from a single scalar or a $d$-vector to a $d \times d$ matrix.
Specifically, the attention weight function in our matrix attention is $\mathcal{W}(\mathbf{f}_i, \mathbf{f}_{n_j}) = \boldsymbol{\Psi}(\mathbf{f}_i - \mathbf{f}_{n_j})$, where $\boldsymbol{\Psi}: \mathbb{R}^{d} \to \mathbb{R}^{d \times d}$ is learnable function implemented as an MLP:
\begin{equation}
\label{eqn:matrix-att}
  \mathbf{z}_i = \sum_{j=1}^{k}\mathtt{Norm}(\boldsymbol{\Psi}(\mathbf{f}_i-\mathbf{f}_{n_j})) \mathbf{f}_{n_j}.
\end{equation}
$\boldsymbol{\Psi}(\mathbf{f}_i-\mathbf{f}_{n_j})$ generates a $d^2$ vector first and we reshape it as a $d\times d$ matrix. In the matrix attention $\boldsymbol{\Psi}(\mathbf{f}_i-\mathbf{f}_{n_j})$, the row of the matrix is channel dimension. With an attention weight matrix, we extend the scalar product in previous two attentions to a matrix product, thus enabling inter-channel feature exchange in the feature aggregation function.
Fig.~\ref{fig:nv} illustrates the comparison of our matrix attention and previous attentions.
Our method can aggregate features from the feature dimension and the spatial dimension simultaneously, which brings more expressive representation for the point cloud.


\paragraph{Normalized Matrix Attention}\label{sec:normalization}

The $\mathtt{Norm}$ function in attention confines the magnitude of the attention weights, which stabilizes the training process. Softmax is a commonly used normalization operation in scaled dot-product attention and vector attention, which focuses on the largest weights. However, it has two drawbacks which will be mentioned below. For this reason, we propose to leverage a linear normalization method in the matrix attention:
\begin{equation}
\label{eqn:linear-norm}
\mathtt{Norm}(\boldsymbol{\Psi}(\mathbf{f}_i - \mathbf{f}_{n_j})) = \frac{\boldsymbol{\Psi}(\mathbf{f}_i - \mathbf{f}_{n_j})}{\sum_{c=1}^{d} \lvert \boldsymbol{\Psi}(\mathbf{f}_i - \mathbf{f}_{n_j})_{\cdot c}\rvert}.
\end{equation}
The normalization is performed on the channel dimension of the attention weight matrix.
Compared with softmax, the linear normalization provides a smoother distribution of the attention weights and the gradient not close to $0$, which can enrich the learned features. 

We can analyze it theoretically. We first write down the two normalization term in Eq.~\ref{eq:softmax} and Eq.~\ref{eq:onorm}.
\begin{equation}
    \label{eq:softmax}
    f(x_i) = \frac{e^{x_i}}{\sum_j e^{x_j}},
\end{equation}

\begin{equation}
    \label{eq:onorm}
    g(x_i) = \frac{x_i}{\sum_j |x_j|}.
\end{equation}
Then, we can obtain
\begin{equation}
    \label{eq:softmax_grad}
    \frac{\partial f(x_i)}{\partial x_i} = f(x_i)(1-|f(x_i)|),
\end{equation}
\begin{equation}
    \label{eq:onorm_grad}
    \frac{\partial g(x_i)}{\partial x_i} = \frac{1}{\sum_j|x_j|}(1-|g(x_i)|).
\end{equation}

We can compare Eq.~\ref{eq:softmax_grad} and Eq.~\ref{eq:onorm_grad}. They have the common term $1-f(x_i)$. When $i$ varies, $\frac{1}{\sum_j|x_j|}$ is a constant term, but $f(x_i)$ is a term which focus on only the largest terms. 

From the Eq.\ref{eq:softmax_grad} and Eq.\ref{eq:onorm_grad}, we can obtain two conclusions, also the two drawbacks of softmax mentioned above. 
\begin{itemize}
    \item The gradient of $g(x_i)$ is more smooth vector than $f(x_i)$ when $i$ varies. $g(x_i)$ can back-propagate to almost all the input, while $f(x_i)$ can only back-propagate to several largest terms. 
    \item the gradient of $f(x_i)$ is prone to vanish. Because $f(x_i)$ is a softmax term, $f(x_i) \rightarrow 0$ or $f(x_i) \rightarrow 1$, then $\frac{\partial f(x_i)}{\partial x_i} \rightarrow 0$. However, $\frac{\partial g(x_i)}{\partial x_i}$ do not have the problem. This is similar to the relation between sigmoid activation and ReLU activation.
\end{itemize} 
we will provide experiments to prove the two conclusion as well in Sec.~\ref{sec:grad_exp}

And the \emph{normalized matrix attention} is computed as:
\begin{equation}
\label{eqn:Tensorformer}
  \mathbf{z}_i =  \sum_{j=1}^{k} \frac{1}{S_j} \boldsymbol{\Psi}(\mathbf{f}_i-\mathbf{f}_{n_j}) \mathbf{f}_{i},
\end{equation}
where $S_j = \sum_{c=1}^{d} \lvert \boldsymbol{\Psi}(\mathbf{f}_i - \mathbf{f}_{n_l})_{\cdot c}\rvert$ is the normalization term.
By injecting the normalized matrix attention, we design the Tensorformer block by adding additional linear layers. The structure of the Tensorformer block is shown in Fig.~\ref{fig:network} (bottom-left).

\subsection{Indicator Module}\label{sec:indicator}

Given the per-point features for the input point cloud, we further design a \emph{Indicator Module} to predict the occupancy value of an arbitrary query point in the space for mesh reconstruction.
To generate the query points, we first voxelize the whole space and sample one query point from each voxel, leading to a set of uniformly scattered query points.

To extract the query point features, previous works \cite{2019DeepSDF,2020Points2Surf} simply concatenates the position of the query point and the features of its neighbors in point cloud. However, this design ignores the spatial relationships between the query point and its neighbors, leading to inferior reconstruction performance. 
To address this issue, we design a novel position-aware attention layer to learn the features for the query points from the features of the input point cloud. Specifically, given a query point $\textbf{q}_i$ and its $k$ nearest neighbors in the input point cloud $\{ \textbf{p}_{n_1}, ..., \textbf{p}_{n_k} \}$ with the features $\{ \textbf{f}_{n_1}, ..., \textbf{f}_{n_k} \}$, the features $\mathbf{g}_i$ of $\mathbf{q}_i$ are computed as:
\begin{equation}
\label{eqn:eq6}
  \mathbf{g}_i = \sum_{j=1}^{k}\boldsymbol{\Omega}(\mathbf{q}_i-\mathbf{p}_{n_j}) \odot \mathbf{f}_{n_j},
\end{equation}
where $\boldsymbol{\Omega}: \mathbb{R}^{3} \to \mathbb{R}^{d}$ is a leanable function implemented as an MLP, and $\odot$ denotes Hadamard product.

After obtaining the per-point features of the query points, another shared MLP $\boldsymbol{\Theta}: \mathbb{R}^d \rightarrow \mathbb{R}$ is adopted to predict the occupancy of the query points.
\begin{equation}
    o_i = \mathtt{Sigmoid}(\boldsymbol{\Theta}(\mathbf{g}_i)).
\end{equation}
At last, we apply Marching Cube \cite{lorensen1987marching} on the predicted occupancy to generate a watertight mesh which is then smoothed with Laplacian smoothing.

To supervise the network training, we use a binary cross entropy loss on the predicted occupancy of the query points:
\begin{equation}\label{eqn:eq7}
L =-\frac{1}{N} \sum_{i=1}^{N} y_{i} \cdot \log \left(o_{i}\right)+\left(1-y_{i}\right) \cdot \log \left(1-o_i\right),
\end{equation}
where $y_{i}$ is the ground-truth occupancy.

\subsection{Discussions}\label{sec:disscuss}

In this subsection, we discuss the differences between point-based convolution, vector attention + FC and our matrix attention.


\paragraph{Matrix Attention vs. Point-based Convolution}
Point-based convolution is another popular line of operators for point cloud feature learning.
In point-based convolution, each neighbor point $\mathbf{p}_{n_j}$ is assigned a group of convolutional weights $\mathcal{W}(\mathbf{p}_{n_j}, \mathbf{p}_i) \in \mathbb{R}^{d \times d}$ according to its \emph{spatial position}. The weights can be fixed (i.e., voxel-based convolution  \cite{choy20194d,graham20183d}), interpolated from a pre-defined weights bank (i.e., KPConv \cite{thomas2019kpconv}, PAConv \cite{xu2021paconv}), or dynamically learned (i.e., PointConv \cite{2019PointConv}). And the output features of the anchor point are computed as a summed matrix multiplication between the features and the convolutional weights of the neighbor points:
\begin{equation}
\mathbf{z}_i = \sum_{j=1}^{k} \mathcal{W}(\mathbf{p}_{n_j}, \mathbf{p}_i) \mathbf{f}_{n_j}.
\end{equation}
Similar with our Tensorformer, feature aggregation in convolution is also conducted in both the spatial dimension and the feature dimension. However, the attention weights in matrix attention are computed from feature similarity, while the convolution weights are computed from spatial positions. This difference allows Tensorformer to learn point features in a more flexible manner, which contributes to stronger expressivity. We compare these two methods in Sec.~\ref{sec:ablation}.



\paragraph{Matrix Attention vs. Vector Attention + FC}
In matrix attention, features in different channels and different neighbor points can be weighted and fused simultaneously. Although information in vector attention cannot flow across channels, adding a fully connected (FC) layer does achieve that. At this point, vector attention + FC can be regarded as a decomposition of matrix attention which is comprised of an intra-channel feature fusion step and an inter-channel feature fusion step.
Fig.~\ref{fig:diff} illustrates the comparison between these two structures.
Although the combination of two achieves feature aggregation in both the spatial dimension and the feature dimension, the information loss in both steps induced by the decomposition cannot be fully compensated.
A similar case here is the comparison between depthwise separable convolution \cite{2017MobileNets} and the regular convolution in 2D convolutional networks, where the former is more lightweight but the latter is more expressive. We also provide comparison between these two structures in Sec.~\ref{sec:ablation}.

\paragraph{Decoupling vs. Coupling}
Decoupling is effective in image understanding especially for convolution network. The standard convolution involves spatial and channel information exchange at the same time. Decoupling can reduce the computation cost but lose minor performance. Its performance is usually weaker than standard convolution. However, standard transformer just aggregates information in spatial dimension not in channel dimension, and information exchange along channel dimension is achieved by FC rather than transformer, sum is only performed in neighbor dimension. Feature in the same channel can be aggregated while feature in different channel do not interact with each other, like in the following equation:
$\sum_{i \in N_i}(w_iv_i)$, where $w_i  = softmax(q_i^T k)$, where $v_i$ is the value in transformer and $q_i$, $k$ are the query and key, $N_i$ is the neighborhood of the query point.

Partially based on this idea, we want to expand the standard transformer so that it, like standard convolution, can aggregate spatial and channel information at the same time. 

\paragraph{Time Complexity}
Here, we write the channel dimension as $d$ and the number of neighbor point as $k$. Then, the time complexity of standard transformer is $O(kd)$, the time complexity of vector attention is $O(kd)$, the time complexity of tensorformer is $O(kd^2)$. Further more, we have provided the time comparison in Tab.~\ref{tab:cost}. We can see that our method need more computation cost than others.

\section{Experiments}\label{4}

In this section, we evaluate the efficacy of our Tensorformer on the 2 datates, including ShapeNetCore \cite{chang2015shapenet}, ABC\cite{abc}. We first present quantitative and qualitative comparison with previous state-of-the-art methods in Sec.~\ref{sec:comparison-sota}. And we conduct comprehensive ablation studies in Sec.~\ref{sec:ablation}.

\paragraph{Metrics} We adopt 3 mainstream metrics for point cloud reconstruction: (1) Chamfer distance ($\mathrm{CD}_1$) which measures the difference between the point cloud sampled from ground-truth shape and predicted shape, (2) Normal Consistency ($\mathrm{NC}$) which measures the difference between ground-truth surface normal and predicted surface normal, and (3) Intersect over Union ($\mathrm{IoU}$) which measures the overlap between ground-truth shape and predicted shape. Please refer to Appendix.~\ref{app:eval} for more details.


\subsection{Implementation Details}
\label{sec:implementation}


\paragraph{Network architecture} Our reconstruction network contains four Tensorformer blocks, with the output dimensions of $8$, $32$, $32$, $32$ and $32$, respectively. We use $k=24$ nearest neighbor points as the local patch. In the indicator module, we first increase the number of channels to $128$, and the final MLP consists of two layers with the feature dimensions of [$32$, $1$]. As for down-sample layer, we adopt farthest point sampling to convert the number of points from 3000 to 512. In order to obtain the feature in down-sampled point, we adopt a learnable method, called Indicator Layer, same as the final prediction layer.
For the query points, we adopt slightly different sampling strategies during training and testing.
During training, we voxelize the space into $64^3$ voxels and apply corrosion and expansion on the voxels to obtain points around the surface. Meanwhile, in order to get the points evenly scattered in the space, we again partition the space into $16^3$ voxels and randomly sample a point from each voxel. This strategy puts more emphasis on the vicinity of the surface while also making the sampled points scattered evenly in the space.
During testing, we simply voxelize the space into $64^3$ voxels and collect the center points of all the voxels as the query points.

\paragraph{Training and testing} Following common practice \cite{2019Occupancy,2020Convolutional,deepmls}, we train and test Tensorformer with the 3D models from 13 shape classes in ShapeNetCore. The models are trained using Adam~\cite{adam} optimizer for 4000000 iterations with a batch size of 2. We use the cosine learning rate decay \cite{2016SGDR} with initial learning rate of $0.0001$. For each shape, we randomly sample 3,000 points and add Gaussian noise with a standard deviation of $0.005$ of the maximum bounding-box side length of the shape. Besides, in comparison experiments, we provide ground-truth normal with a Gaussian noise with a standard deviation of $0.1$ for PSR and BPA. As for the data augmentation, we random flip each point cloud along a random axis (x-, y- or z-axis) to prevent the network from memorizing fixed shapes.


\subsection{Gradient comparison between softmax normalization and our normalization}\label{sec:grad_exp}
\begin{figure} 
\centering
  \includegraphics[width=\columnwidth, height=0.35\columnwidth]{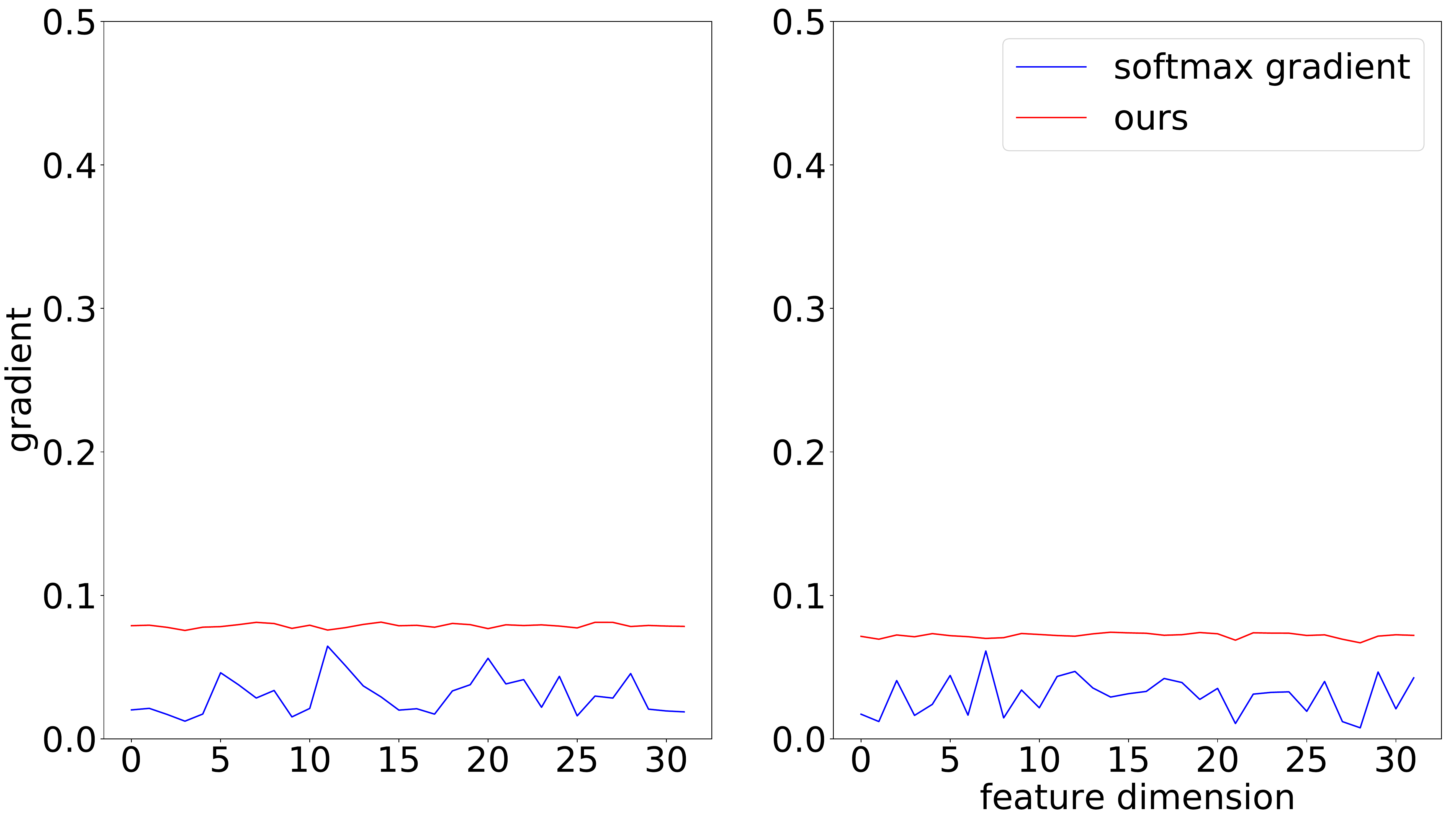}
  \caption{The gradient comparison between softmax normalization and ours.}
  \label{fig:grad}
\end{figure}

We extract the gradient of Softmax normalization and ours from the first Tensorformer layer. Then, we show gradient result in Fig.~\ref{fig:grad}. We can see that the red line is smoother than the blue line, and the red line is far from $0$ which means gradient vanishing. Those supports the theoretical in Sec.~\ref{sec:normalization}.

\subsection{Surface reconstruction from point cloud on ABC}
In order to show the priority of Tensorformer, we first show the result on ABC dataset. In ABC dataset, we only choose all the 7440 meshes from 001. Then we randomly partition 30\% meshes for testing, the remaining 70\% for training. We compare our method on ABC dataset with four methods, ConvOcc\cite{2020Convolutional}, Points2Surf\cite{2020Points2Surf}, PSR\cite{2013Poisson} and BPA\cite{ballpivoting}. The quantitative result can be found in Tab.~\ref{tab:abc}. we can see that our method achieves best result on all metrics. 

\begin{figure*} 
\centering
  \includegraphics[width=1.8\columnwidth]{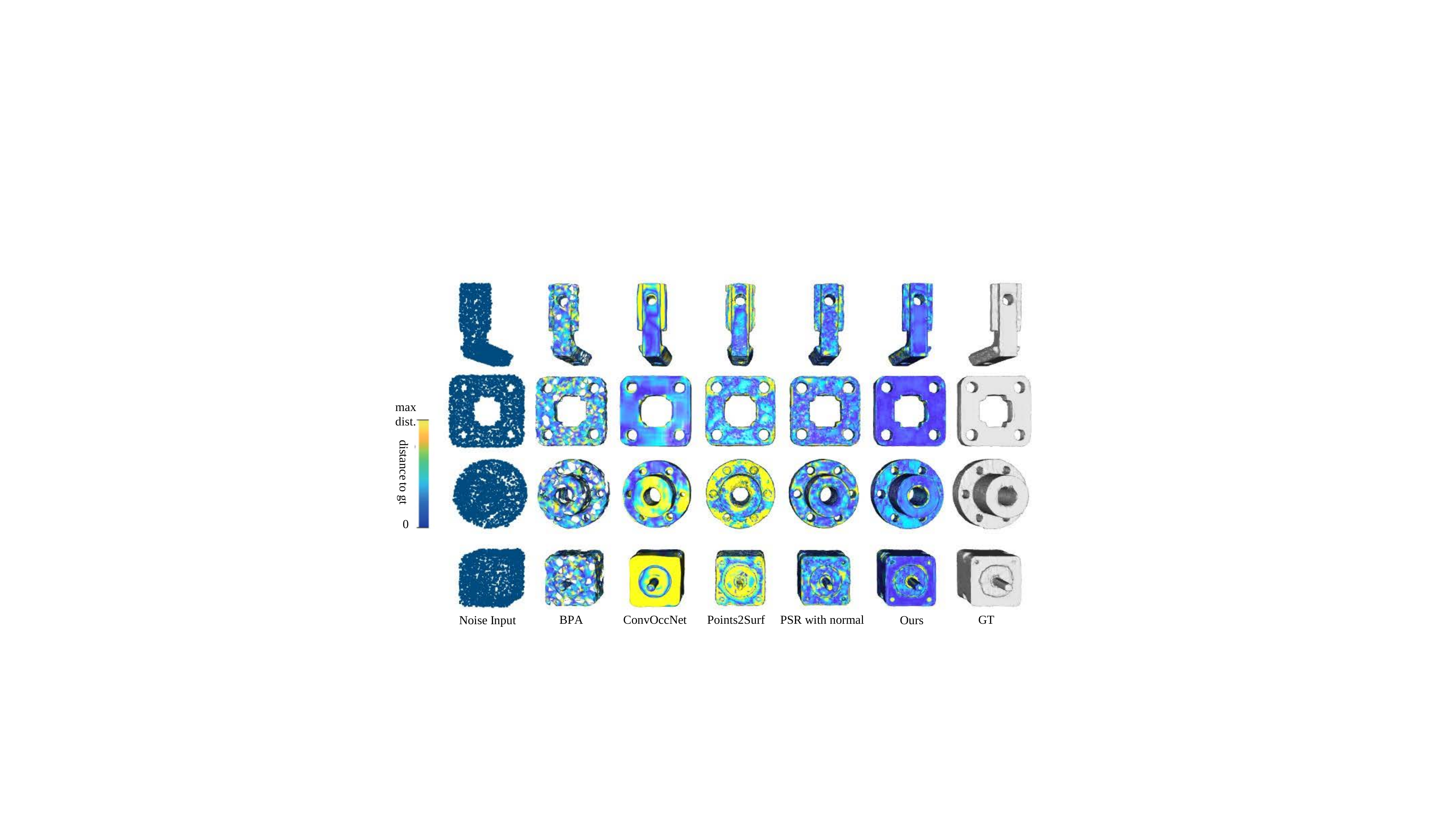}
  \caption{Qualitative results of Tensorformer and the baseline methods on ABC.}
  \label{fig:abc_vis}
\end{figure*}

Further more, we provide several visualization result in Fig.~\ref{fig:abc_vis}. Obviously, the mesh reconstructed from our method is closer to the ground-truth mesh because our mesh contains more blue area. Our method can generate more smooth and faithful surface than others. More visualization results can be found in Appendix.~\ref{app:vis}.
\begin{table}[]
\centering
\caption{Point cloud reconstruction results on ABC dataset.}
\begin{tabular}{cccc}
\toprule
Method & {$\mathrm{CD}_{1} \downarrow$} & {$\mathrm{NC} \uparrow$} & {$\mathrm{IoU} \uparrow$}\\
\midrule
Points2Surf\cite{2020Points2Surf} & 0.0668 & 0.8771 & 0.7338\\
ConvOccNet~\cite{2020Convolutional} & 0.0599 & 0.9396& 0.8394\\
BPA\cite{ballpivoting} & 0.0802 &  0.8609& n/a\\
PSR\cite{2013Poisson} & 0.0366 &  0.9426 & 0.9159\\
{Tensorformer (Ours)} & \textbf{0.03635} & \textbf{0.9485}& \textbf{0.9164}\\
\bottomrule
\end{tabular}
\label{tab:abc}
\end{table}

\subsection{Surface reconstruction from point cloud on ShapeNet}\label{sec:comparison-sota}

\begin{table}[]
\centering
\caption{Point cloud reconstruction results averaged over the 13 categories of ShapeNetCore.}
\begin{tabular}{cccc}
\toprule
Method & {$\mathrm{CD}_{1} \downarrow$} & {$\mathrm{NC} \uparrow$} & {$\mathrm{IoU} \uparrow$}\\
\midrule
O-CNN-C~\cite{o-cnn} & 0.0606 & 0.9414  & n/a\\
ConvOccNet~\cite{2020Convolutional} & 0.0405 & 0.9423  &0.9065\\
IMLSNet~\cite{deepmls} & 0.0301 &  0.9440 & 0.9224\\
PSR\cite{2013Poisson} & 0.0367 &  0.9398 & 0.9060\\
{Tensorformer (Ours)} & \textbf{0.0276} & \textbf{0.9630} &\textbf{0.9622}\\
\bottomrule
\end{tabular}
\label{tab:main_result}
\end{table}

\begin{figure*} 
\centering
  \includegraphics[width=2\columnwidth]{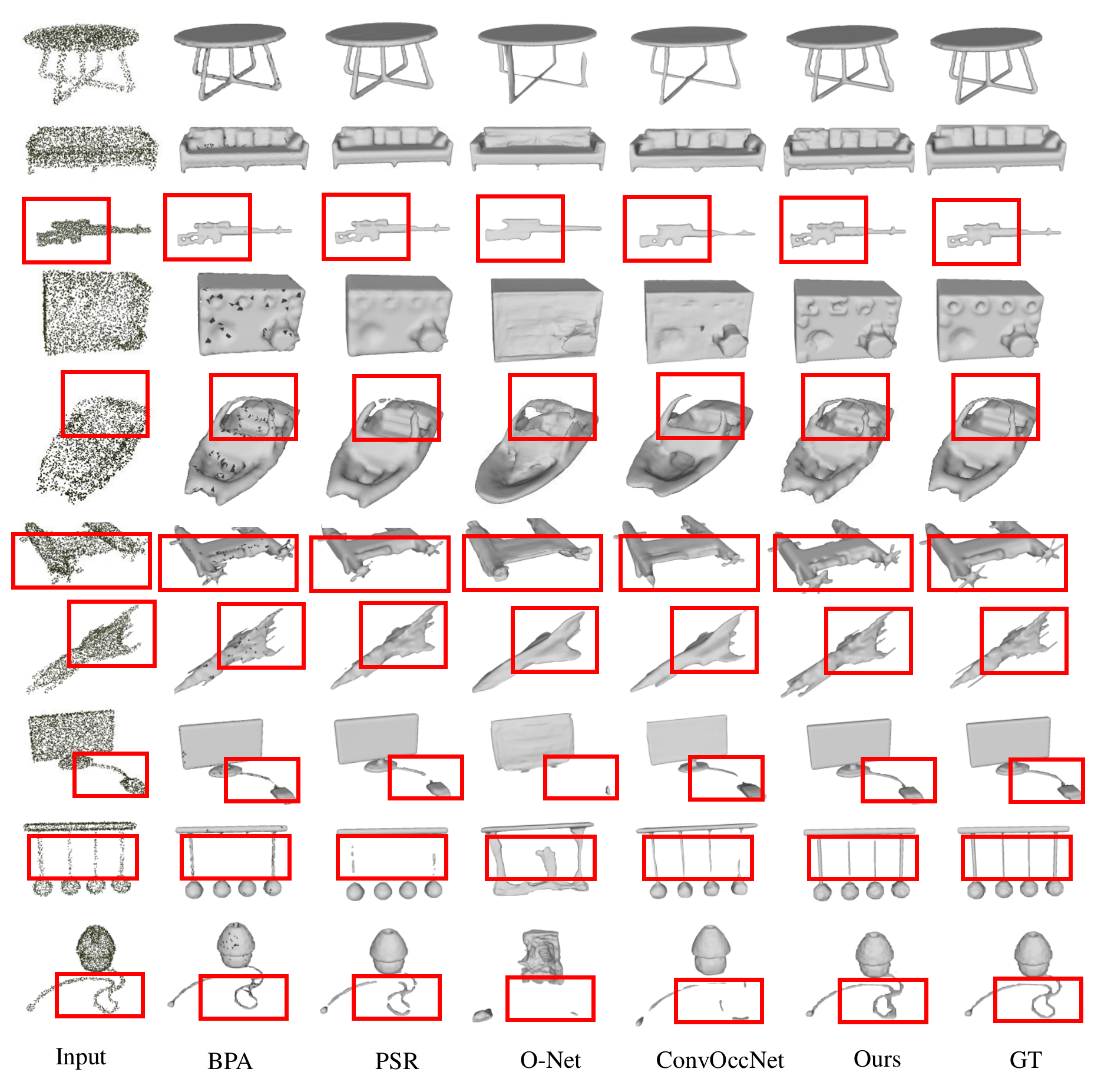}
  \caption{Qualitative results of Tensorformer and the baseline methods. Tensorformer achieves significantly better reconstruction of thin structures.}
  \label{fig:main_result}
\end{figure*}

\paragraph{Quantitative results}
We compare our method with three recent state-of-the-art methods: O-CNN-C \cite{o-cnn}, Convolutional Occupancy Network (ConvOccNet) \cite{2020Convolutional} and IMLSNet \cite{deepmls}. Because ConvOccNet\cite{2020Convolutional}is the successive method of O-Net\cite{2019Occupancy} and ConvOccNet\cite{2020Convolutional} presents that ConvOccNet performs much better than both
O-Net\cite{2019Occupancy} and DeepSDF\cite{2019DeepSDF}, we do not compare our method with DeepSDF and O-Net.
As shown in Tab.~\ref{tab:main_result}, Tensorformer outperforms the baselines on all the three metrics. It is noteworthy that our method achieves a significant improvement on the IoU, demonstrating that the network can predict more accurate occupancy values. We further provide the per-class comparison results in Appendix.~\ref{app:number}. Our method achieve best result on most classes.



\paragraph{Qualitative results}

We also compare the qualitative results of Tensorformer with two traditional methods (Poisson Surface Reconstruction (PSR)\cite{2013Poisson} and Ball Pivoting Algorithm (BPA)\cite{ballpivoting}) and two learning-based methods, O-Net\cite{o-cnn} and ConvOccNet\cite{2020Convolutional}.
As shown in Fig.~\ref{fig:main_result}, our method performs significantly better in reconstructing thin structures (see the fans in the $1^{\text{st}}$ row and the wires in the last three rows). Note that BPA and PSR use the ground-truth normals as input while our method only uses the point clouds. Compared with the learning-based O-Net and ConvOccNet, Tensorformer can preserve more fine-grained geometric structures and produce more detailed reconstructions. There are 3 reasons of good performance. First, Tensorformer can capture better local geometry due to the stronger expression ability. Second, the point-aware indicator layer not only considers the relative distance from query point to point cloud but also considers the direction of the relative shift, intuitively, relative shift from point cloud to query point is important for occupancy value classification. Thirdly, the supervision information is also important, because we sample the query points around the surface, our network can learn the thin structure from training data.


\paragraph{Robustness analysis}

We further evaluate the robustness of Tensorformer to the variation of the number of sampled points. For simplicity, we only report results on the plane class. As shown in Tab.~\ref{tab:nsample}, our method achieves consistently better performance as more points are used, but the improvements are smaller as the more points are used. These results demonstrate the strong robustness of our method to different numbers of sampled points.
Unless otherwise noted, we sample $3000$ points in all experiments.

We also investigate the influence of the number of local neighbor points in the attention block. We increase the number of neighbor points $k$ from $12$ to $48$ in Tab.~\ref{tab:knn}. It can be observed that Tensorformer is quite stable at least when $k$ varies in this range.


\begin{table}[]
\centering
\caption{Point cloud reconstruction results on the airplane class of ShapeNetCore when the number of sample points varies.}
\begin{tabular}{cccc}
\toprule
\# Points & {$\mathrm{CD}_{1} \downarrow$} & {$\mathrm{NC} \uparrow$} & {$\mathrm{IoU} \uparrow$}\\
\midrule
1024 & 0.0309 & 0.9254 &0.9216\\
2048 & 0.0257 & 0.9443 &0.9449\\
3000 &0.0219  & 0.9558 &0.9582\\
4096 & \textbf{0.0207} & \textbf{0.9626} &\textbf{0.9647}\\
\bottomrule
\end{tabular}
\label{tab:nsample}
\end{table}

\begin{table}[]
\centering
\caption{Point cloud reconstruction results on the airplane class of ShapeNetCore when the number of neighbor points varies, where $k$ is the number of neighbor points.}
\begin{tabular}{cccc}
\toprule
$k$ & {$\mathrm{CD}_{1} \downarrow$} & {$\mathrm{NC} \uparrow$} & {$\mathrm{IoU} \uparrow$}\\
\midrule
12 &  0.0242& 0.9479 &0.9492\\
24 & 0.0219 & 0.9558 &0.9582\\
36 &  \textbf{0.0213}& \textbf{0.9573}&\textbf{0.9603}\\
48 & 0.0216 & 0.9566 &0.9593\\
\bottomrule
\end{tabular}
\label{tab:knn}
\end{table}

\paragraph{Generalization analysis}

\begin{figure*} 
\centering
  \includegraphics[width=1.85\columnwidth]{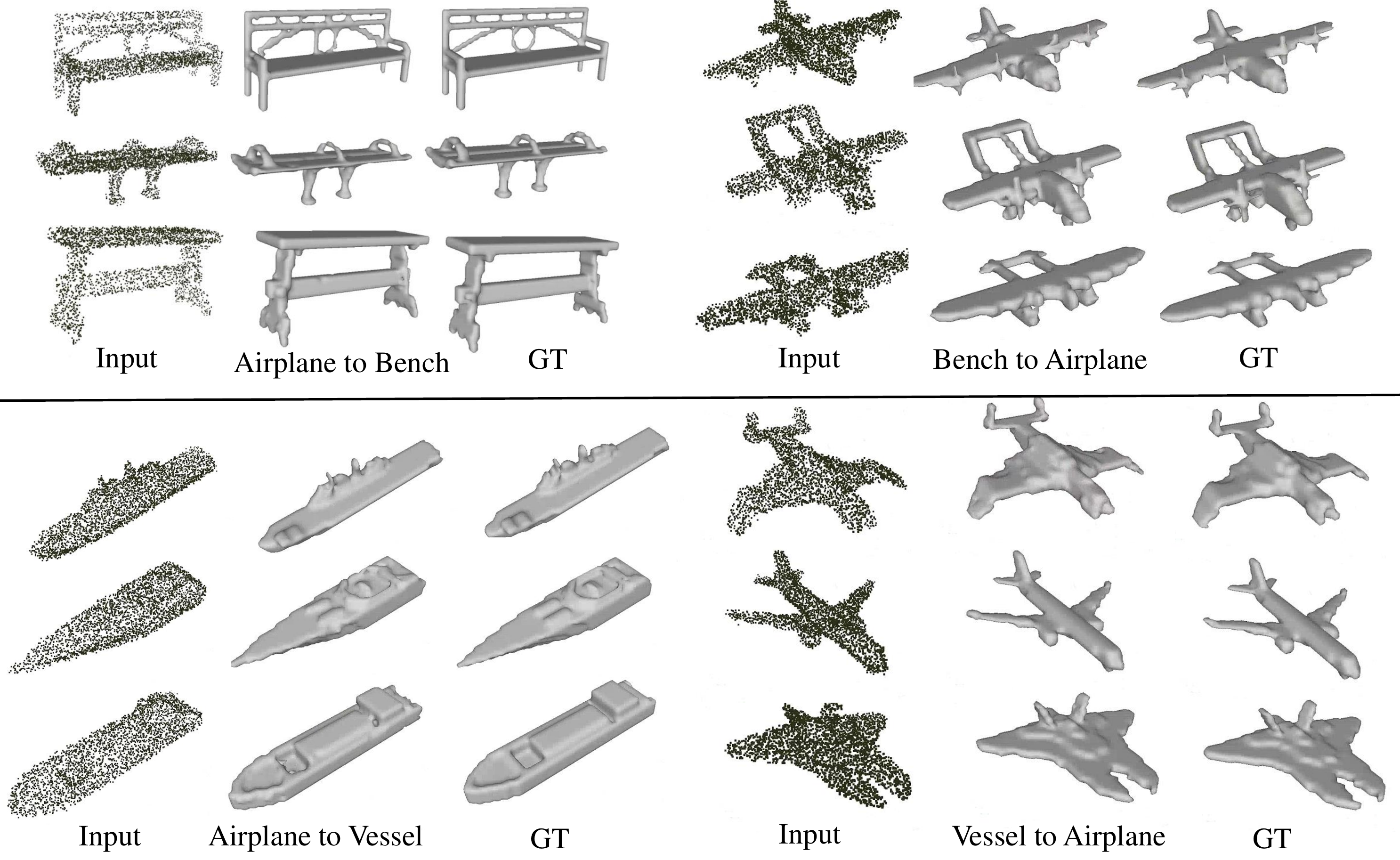}
  \caption{Generalization results. ``A to B'' means training on class A and testing on class B. GT is ground-truth.}
  \label{fig:g1}
\end{figure*}

To investigate the generality to unseen categories, we evaluate the performance of Tensorformer by training and testing on different categories. As shown in Tab.~\ref{tab:generalization}, the models under this setting achieve very close performance to the models trained and tested on the same categories. As our indicator layer only uses local information to reconstruct the mesh, our method can capture more general and repeatable local geometries across different shape categories, thus contributing to good generality. Fig.~\ref{fig:g1} illustrates some generalization reconstruction results.

\begin{table}[]
\centering
\caption{Generalization results on ShapeNetCore.}
\begin{tabular}{lllll}
\toprule
Train    & Test     & \textbf{$\mathrm{CD}_{1} \downarrow$} & \textbf{$\mathrm{NC} \uparrow$} & \textbf{$\mathrm{IoU} \uparrow$} \\
\midrule
vessel   & airplane & 0.0227 & 0.9635 & 0.9538  \\  
bench    & airplane & 0.0603 & 0.9166 & 0.8691  \\ 
airplane & airplane   & 0.0183  & 0.9694 & 0.9645  \\ 
\midrule
airplane & vessel   & 0.0378  & 0.9243 & 0.8963  \\
vessel   & vessel & 0.0229 & 0.9522 & 0.9525  \\ 
\midrule
airplane & bench    & 0.0245    & 0.9611    & 0.9395 \\ 
bench    & bench & 0.0249 & 0.9588 & 0.9339  \\
\bottomrule
\end{tabular}
\label{tab:generalization}
\end{table}

\subsection{Ablation Study}
\label{sec:ablation}

In this subsection, we conduct ablation studies to provide a better understanding of the effectiveness of different modules in Tensorformer.

\paragraph{Experiment setting} We compare six models to learn the point-wise features: (a) PointConv \cite{2019PointConv}, (b) scaled dot-product attention \cite{zhao2021pointtrans}, (c) vector attention + FC \cite{2020Vector}, (d) matrix attention with softmax normalization, (e) matrix attention without normalization and (f) matrix attention with normalization (Tensorformer). We evaluate the networks using $3$ categories (i.e., vessel, bench and airplane). All models adopt the same overall architecture as in Fig.~\ref{fig:network}. And we carefully tune the number of feature channels in the models to guarantee they take similar run-time memory footprint.

Although achieving good performance, one drawback of our method is longer time and larger memory cost. We list the time and memory cost in Tab.~\ref{tab:cost}. We run 100 random reconstruction examples and average the time and memory cost. We can see that the time and memory cost is indeed larger than the baseline method, but the cost do not increase too much and is acceptable. 

\begin{table}[]
\caption{Time and memory cost comparison.}
\label{tab:cost}
\begin{tabular}{lll}
\toprule
                             & memory footprint(MB) & inference time(s) \\ \midrule

Scaled Dot-product Attention & 1679 & 3.54 \\ 
Vector Attention             & 1219                 & 3.53 \\
Point Transformer            & 1290                 & 4.23      \\
Tensorformer  & \textbf{2710}  & \textbf{5.23}        \\
\bottomrule
\end{tabular}
\end{table}



\paragraph{Results and analysis}
Tab.~\ref{tab:ab} shows the results of the ablation experiments. The models based on matrix attention outperforms the models with scaled dot-product attention and vector attention significantly. As matrix attention leverages inter-channel information communication during feature aggregation, the resultant features are more expressive and better models local geometric structures.
This is consistent with our insight to design matrix attention.
And the model with the proposed linear normalization surpasses the models with softmax and without normalization, which demonstrates the effectiveness of this design. It is noteworthy that the model without normalization performs better than the softmax-based one, indicating that softmax cannot handle the large attention weights matrix in matrix attention. Moreover, the performance of point convolution is inferior to most attention-based methods, which shows that attention is a more effective method to learn feature representations for point clouds.

\begin{table}[]
\centering
\caption{Comparison with the basic transformer layer.}
\label{tab:ab}
\begin{tabular}{lllllll}
\toprule
Method & $\mathrm{CD}_1 \downarrow$ & NC$\uparrow$ & IoU$\downarrow$ \\
\toprule
(a) Scaled Dot-Product Attention & $0.0438$ & $0.9337$ & $0.9316$ \\
(b) Vector Attention + FC & $0.0513$ & $0.8840$ & $0.8735$ \\
(c) Matrix Attention + Softmax & $0.0422$ & $0.9424$ & $0.9412$ \\
(d) Matrix Attention w/o Norm & $0.0369$ & $0.9510$ & $0.9503$ \\
(f) PointConv & $0.0477$ & $0.9413$ & $0.9395$ \\
\midrule
(e)Tensorformer & $0.0276$ & $0.9630$ & $0.9622$ \\
\bottomrule
\end{tabular}
\end{table}

\section{Conclusion}
We have presented Tensorformer based on two key mechanisms, matrix attention and weight normalization. We apply it to point cloud reconstruction and achieve high quality result. 
The main limitation of our current method is the relatively high computation complexity which is $O(kd^2)$ with $d$ being feature dimension and $k$ the number of neighbor points. Therefore, when we expand the network in width, the memory and computational cost will increase rapidly. In the future, we plan to optimize it with random mechanism or matrix attention decomposition, then apply it to classification and segmentation task.

\section{Acknowledgements}
This research is funded by the following projects. National Key Research and Development Program of China(2018AAA0102200). National Natural Science Foundation of China(62002375, 62002376, 62132021, 62325211, 62372457). National Science Foundation of Hunan Province of China(2021JJ40696, 2021RC3071, 2022RC1104). NUDT Research Grants(ZK22-52).

\clearpage
\bibliographystyle{IEEETran}
\bibliography{bibfile}

\appendix

\section{More Visualization Result}
\label{app:vis}
\begin{figure*} 
\centering
  \includegraphics[width=2\columnwidth]{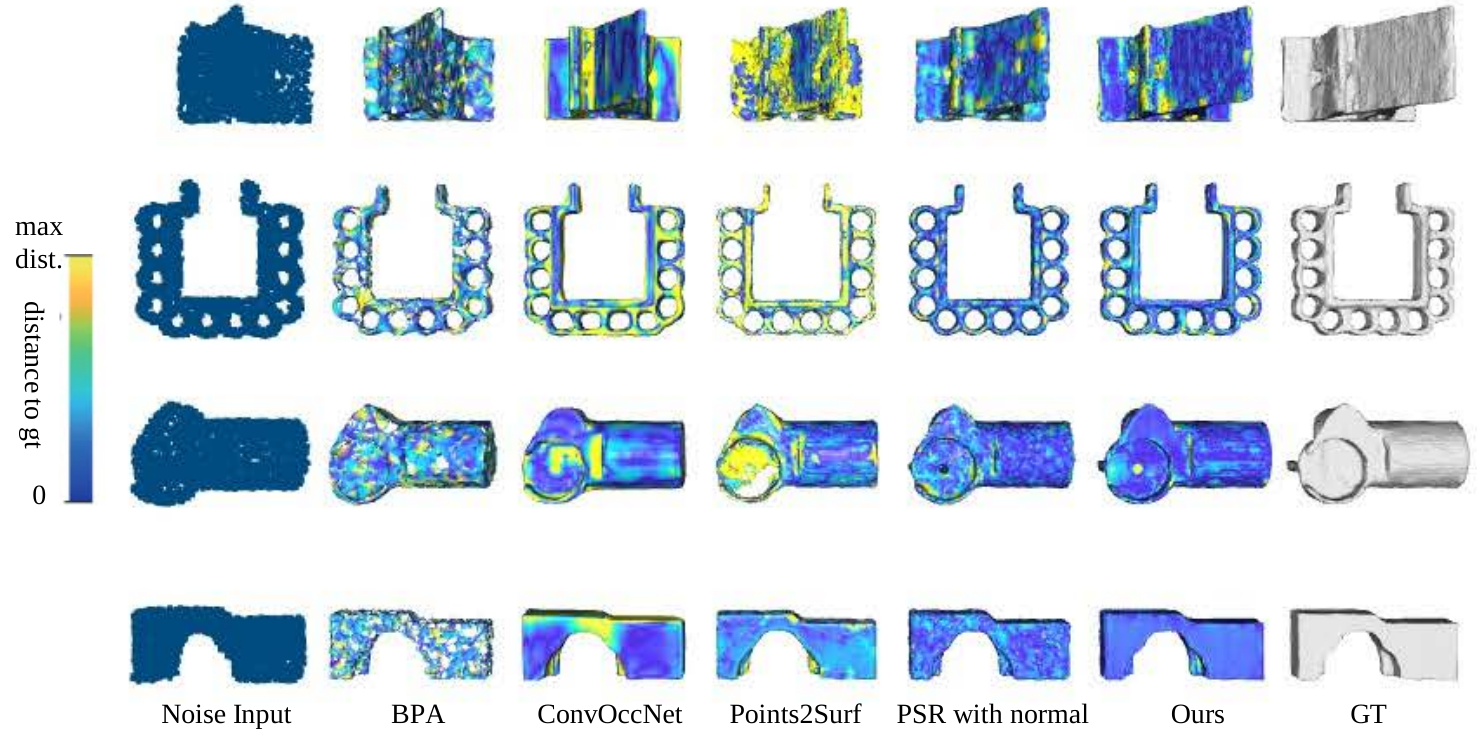}
  \caption{Qualitative results of Tensorformer and the baseline methods on ABC.}
  \label{fig:abc_vis2}
\end{figure*}

\begin{figure*} 
\centering
  \includegraphics[width=2\columnwidth]{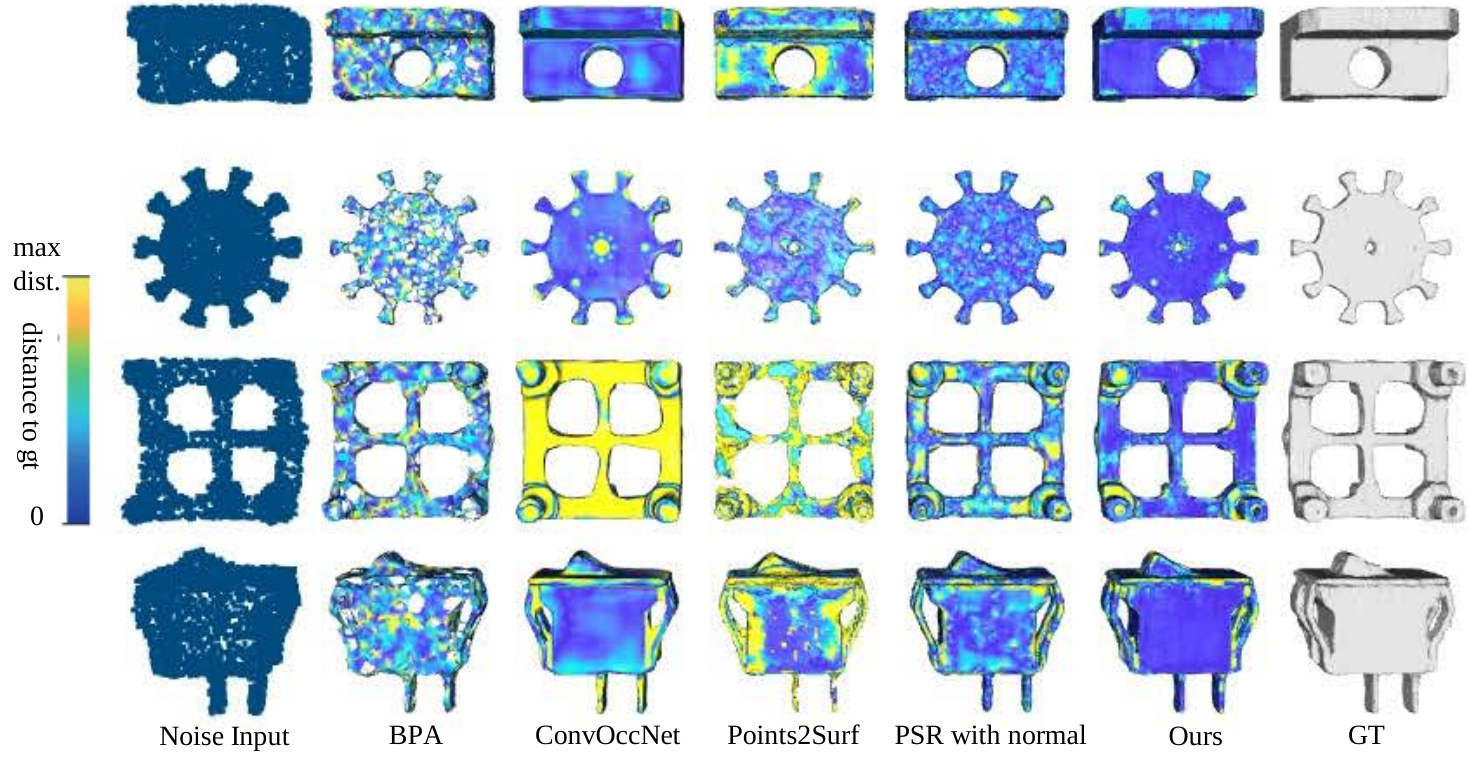}
  \caption{Qualitative results of Tensorformer and the baseline methods on ABC.}
  \label{fig:abc_vis3}
\end{figure*}

\begin{figure*} 
\centering
  \includegraphics[width=2\columnwidth]{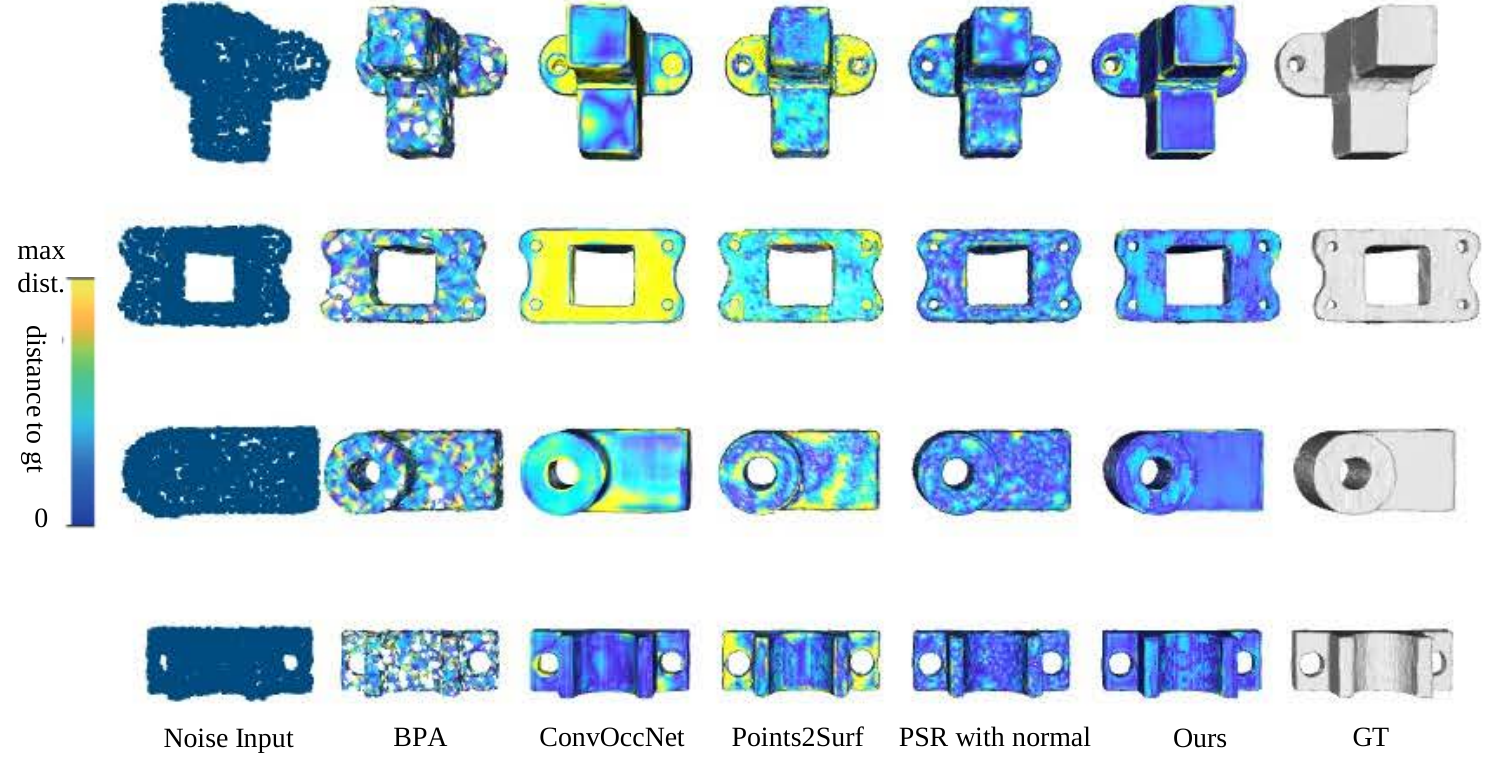}
  \caption{Qualitative results of Tensorformer and the baseline methods on ABC.}
  \label{fig:abc_vis4}
\end{figure*}

More visualization result on ABC dataset can be found in Fig. \ref{fig:abc_vis2}, Fig. \ref{fig:abc_vis3}, Fig. \ref{fig:abc_vis4}.

\section{More Quantitative Result}\label{app:number}
More metric results on ShapeNet are shown in Table. \ref{tab:cd}, Table. \ref{tab:nc} and Table. \ref{tab:iou}.

\begin{table*}[]
\centering
\begin{tabular}{c|ccccc}
\hline & \multicolumn{5}{|c}{$\mathrm{CD}_{1} \downarrow$} \\
Category & O-CNN-C & ConvOccNet & IMLSNet & PSR&  Tensorformer (Ours) \\
\hline airplane & $0.0574$ & $0.0316$ & $0.0240$& $0.0235$ &$\mathbf{0.0183}$\\
bench & $0.0586$ & $0.0312$ & $0.0296$ &$0.0371$& $\mathbf{0.0249}$\\
cabinet & $0.0649$  & $0.0411$ & $0.0343$ &$0.0558$&$\mathbf{0.0327}$\\
car & $0.0705$  & $0.0702$ & $0 . 0 3 9 0$&$0.0393$ & $\mathbf{0.0272}$\\
chair & $0.0604$  & $0.0419$ & $0 . 0 3 4 3$ &$0.0525$&$\mathbf{0.0316}$\\
display & $0.0595$ & $0.0338$ & $0 . 0 2 87$ &$0.0426$&$\mathbf{0.0276}$\\
lamp & $0.0607$ & $0.0535$ & $\mathbf{0 . 0 3 07}$ &$0.0355$& $0.0366$\\
speaker & $0.0669$  & $0.0572$ & $\mathbf{0 . 0 3 9 1}$ &$0.0502$& $0.0508$\\
rifle & $0.0557$ & $0.0240$ & $0 . 0 2 0 2$ &$\mathbf{0.0122}$&$0.0145$\\
sofa & $0.0597$  & $0.0394$ & $0 . 0 3 0 4$ &$0.0378$&$\mathbf{0.0257}$\\
table & $0.0603$  & $0.0345$ & $0 . 0 3 1 4$ &$0.0428$&$\mathbf{0.0299}$\\
telephone & $0.0577$  & $0.0230$ & $0 . 0 2 2 4$ &$0.0239$&$\mathbf{0.0215}$\\
vessel & $0.0581$  & $0.0390$ & $0 . 0267$ &$0.0232$&$\mathbf{0.0229}$\\
\hline mean & $0.0606$ & $0.0405$ & $0.0301$ &$0.0366$&$\mathbf{0.0276}$\\
\hline
\end{tabular}
    \caption{$\mathrm{CD}_{1}$ on ShapeNet on 13 categories.}
    \label{tab:cd}
\end{table*}

\begin{table*}[]
\centering
\begin{tabular}{c|ccccc}
\hline & \multicolumn{5}{|c}{$\mathbf{N C} \uparrow$} \\
Category & O-CNN-C & ConvOccNet & IMLSNet  &PSR& Tensorformer (Ours) \\
\hline airplane& $0.9281$  & $0.9371$ & $0 . 9 3 7 1$&$0.9435$&$\mathbf{0.9694}$ \\
bench & $0.9236$ & $0.9301$ & $0 . 9 2 2 0$ &$0.9062$&$\mathbf{0.9588}$\\
cabinet &$0.9511$ & $0 . 9 5 8 1$ & $0.9546$&$0.9357$&$\mathbf{0.9674}$ \\
car & $0.8768$ & $0 . 8 9 377$ & $0.8820$&$0.8974$&$\mathbf{0.9562}$ \\
chair &$0.9507$ & $0.9466$ & $0 . 9 5 0 3$&$0.9311$&$\mathbf{0.9639}$ \\
display &$0.9698$& $0.9710$ & $0 . 9 7 3 2$&$0.9643$&$\mathbf{0.9787}$ \\
lamp & $0.9211$ & $0.9102$ & $0 . 9 2 1 8$ &$0.9366$&$\mathbf{0.9348}$\\
speaker &$0.9463$  & $0.9445$ & $0 . 9 4 7 3$&$0.9531$&$\mathbf{0.9570}$ \\
rifle &$0.9420$ &  $0.9317$ & $0 . 9 4 3 3$&$\mathbf{0.9685}$&$0.9682$ \\
sofa & $0.9592$ & $0.9622$ & $0 . 9 6 3 1$ &$0.9534$&$\mathbf{0.9791}$\\
table &$0.9561$ &  $0.9613$ & $0 . 9 6 2 1$&$0.9381$&$\mathbf{0.9737}$ \\
telephone &$0.9837$ & $0.9843$ & $0 . 9 8 3 9$&$0.9790$&$\mathbf{0.9842}$ \\
vessel &$0.9321$ & $0.9237$ & $0 . 9 3 1 9$&$0.9464$&$\mathbf{0.9522}$ \\
\hline mean & $0.9414$ & $0.9423$ & $0 . 9 4 4 0$&$0.9426$&$\mathbf{0.9630}$ \\
\hline
\end{tabular}
    \caption{NC on ShapeNet on 13 categories.}
    \label{tab:nc}
\end{table*}

\begin{table*}[]
\centering
\begin{tabular}{c|ccccc}
\hline & \multicolumn{5}{|c}{ IoU $\uparrow$} \\
Category & O-CNN-C & ConvOccNet & IMLSNet&PSR& Tensorformer (Ours) \\
\hline airplane & $\mathrm{n} / \mathrm{a}$ & $0.8613$ & $0.8997$&$0.9156$&$\mathbf{0.9645}$ \\
bench & n/a  & $0.8398$ & $0 . 8 5 8 0$ &$0.8796$&$\mathbf{0.9339}$ \\
cabinet & n/a  & $0.9498$ & $0 . 9 5 9 5$ &$0.9064$&$\mathbf{0.9642}$\\
car & n/a  & $0.8958$ & $0 . 9 1 5 2$ &$0.8722$& $\mathbf{0.9750}$\\
chair & n/a & $0.8809$ & $0 . 9 1 3 3$ &$0.9090$&$\mathbf{0.9553}$\\
display & n/a  & $0.9375$ & $0 . 9 5 9 1$ &$0.9326$&$\mathbf{0.9752}$\\
lamp & n/a & $0.8040$ & $0 . 8 6 8 3$ &$0.9060$&$\mathbf{0.9103}$\\
speaker & n/a  & $0.9312$ & $0 . 9 5 5 0$&$0.9257$&$\mathbf{0.9590}$ \\
rifle & n/a &  $0.8559$ & $0 . 8 9 5 6$ &$0.9426$&$\mathbf{0.9513}$\\
sofa & n/a & $0.9462$ & $0 . 9 6 4 1$ &$0.9351$&$\mathbf{0.9863}$\\
table & n/a & $0.8997$ & $0 . 9 1 6 6$ &$0.9060$&$\mathbf{0.9724}$\\
telephone & n/a & $0.9647$ & $0.9717$ &$0.9511$&$\mathbf{0.9849}$\\
vessel & n/a & $0.8773$ & $0 . 9 2 4 0$&$0.9245$&$\mathbf{0.9525}$ \\
\hline mean & n/a & $0.9065$ & $0 . 9 224$ &$0.9159$&$\mathbf{0.9622}$\\
\hline
\end{tabular}
    \caption{IoU on ShapeNet on 13 categories.}
    \label{tab:iou}
\end{table*}

\section{Evaluation Metric}\label{app:eval}
We adopt 3 metrics to measure the reconstruction quality of our method, chamfer-l1, normal consistency and IoU. F-score is also commonly used metric, but we do not use it. F-score is based on SDF value and we only use in/out information as supervision during training.
When calculating chamfer-l1, we follow O-Net\cite{2019Occupancy} and randomly sample 100000 points on the reconstructed mesh and 100000 points on ground-truth mesh. chamfer-l1 equation is shown as the Eq.\ref{eqn:eq8}
\begin{equation}
\label{eqn:eq8}
\begin{aligned}
\mathrm{CD}_{1}=& \frac{1}{2 N_{x}} \sum_{i=1}^{N_{x}}\left\|\mathbf{x}_{i}-\mathcal{S}_{y }\left(\mathbf{x}_{i}\right)\right\|_1+\\
& \frac{1}{2 N_{y}} \sum_{i=1}^{N_{y}}\left\|\mathbf{y}_{i}-\mathcal{S}_{x}\left(\mathbf{y}_{i}\right)\right\|_1,
\end{aligned}
\end{equation}
$\{ \mathbf{x}_{i}, i=0\cdots N_x\}$ is the points set sampled from ground-truth mesh. $\{\mathbf{y}_{i}, i=0\cdots N_y\}$ is the point set sampled from reconstructed mesh. $\mathcal{S}_{y}(\mathbf{x}_{i})$ means the nearest point to $\mathbf{x}_{i}$ in reconstructed point set. $\mathcal{S}_{x}(\mathbf{y}_{i})$ means the nearest point to $\mathbf{y}_{i}$ in ground-truth point set. $\|\cdot\|_1$ means $L1-distance$.

Normal consistency is closely related to charm-l1, shown in Eq.\ref{eqn:eq9}
\begin{equation}
\label{eqn:eq9}
\begin{aligned}
\mathrm{NC}=& \frac{1}{2 N_{x}} \sum_{i=1}^{N_{x}}<\mathbf{n}\left(\mathbf{x}_{i}) , \mathbf{n}(\mathcal{S}_{y}\left(\mathbf{x}_{i}\right)\right)>+\\
& \frac{1}{2 N_{y}} \sum_{i=1}^{N_{y}}<\mathbf{n}\left(\mathbf{y}_{i}) , \mathbf{n}(\mathcal{S}_{x}\left(\mathbf{y}_{i}\right)\right)> .
\end{aligned}
\end{equation}
$\mathbf{n}(x)$ means the normal of point $x$. $<,>$ means inner-product.

IoU is the quotient of two number, the number of intersection voxel between reconstruction mesh and ground-truth mesh, the number of union voxel between reconstruction mesh and ground-truth mesh. shown in Eq.\ref{eqn:eq10}
\begin{equation}
\label{eqn:eq10}
\begin{aligned}
\mathrm{IOU}=& \frac{|\mathbf{G} \cap \mathbf{R}|}{|\mathbf{G} \cup \mathbf{R}|}.
\end{aligned}
\end{equation}
$\mathbf{G}$ is the set of voxels occupied by ground-truth mesh. $\mathbf{R}$ is the set of voxels occupied by reconstructed mesh. $|\cdot|$ means the cardinality of set. In our experiments, we use detection points rather than the voxels to calculate IoU.
\end{document}